\documentclass[12pt]{article}
\usepackage{amsmath}
\usepackage{graphicx}
\usepackage{enumerate}
\usepackage{natbib}
\usepackage{url} % not crucial - just used below for the URL 

\usepackage{amssymb}
\usepackage{bm}
\usepackage{multirow}

\newcommand{\blind}{1}

\begin{document}

\def\spacingset#1{\renewcommand{\baselinestretch}%
{#1}\small\normalsize} \spacingset{1}

%%%%%%%%%%%%%%%%%%%%%%%%%%%%%%%%%%%%%%%%%%%%%%%%%%%%%%%%%%%%%%%%%%%%%%%%%%%%%%

\if1\blind
{
  \title{\bf Semiparametric Imputation using Conditional Gaussian Mixture Models under Item Nonresponse}
  \author{Danhyang Lee \hspace{.2cm}\\
    Department of Information Systems, Statistics and Management, \\
    University of Alabama\\
    and \\
    Jae Kwang Kim\\
    Department of Statistics, Iowa State University}
  \maketitle
} \fi

\if0\blind
{
  \bigskip
  \bigskip
  \bigskip
  \begin{center}
    {\LARGE\bf Semiparametric Imputation using Conditional Gaussian Mixture Models under Item Nonresponse}
\end{center}
  \medskip
} \fi

\bigskip
\begin{abstract}
Imputation is a popular technique for handling item nonresponse in survey sampling. Parametric imputation is based on a parametric model for imputation and is less robust against the failure of the imputation model. Nonparametric imputation is fully robust but is not applicable when the dimension of covariates is large due to the curse of dimensionality. Semiparametric imputation is another robust imputation based on a flexible model where the number of model parameters can increase with the sample size. In this paper, we propose another semiparametric imputation based on a more flexible model assumption than the Gaussian mixture model. In the proposed mixture model, we assume a conditional Gaussian model for the study variable given the auxiliary variables, but the marginal distribution of the auxiliary variables is not necessarily Gaussian. We show that the proposed mixture model achieves a lower approximation error bound to any unknown target density than the Gaussian mixture model in terms of the Kullback-Leibler divergence. The proposed method is applicable to high dimensional covariate problem by including a penalty function in the conditional log-likelihood function. The proposed method is applied to 2017 Korean Household Income and Expenditure Survey conducted by Statistics Korea. Supplementary material is available online.
\end{abstract}

\noindent%
{\it Keywords:}  Density ratio model, Kullback-Leibler divergence, Survey sampling
\vfill

\newpage
\spacingset{1.5} % DON'T change the spacing!
\section{Introduction}
\label{sec:intro}

Item nonresponse is often encountered in many applications of statistics. Imputation is a popular tool for handling item nonresponse by replacing missing values with a plausible value (or a set of plausible values). Imputation is used to achieve the following goals: Standard data analyses can be applied and the analyses from different users can be consistent. In addition, we make full use of information, leading to more efficient results and may reduce possible nonresponse biases by choosing an appropriate imputation model. \citep{kalton1986treatment}

\cite{rubin1996multiple} proposed multiple imputation (MI) which fills in each missing data with several plausible values to account for full uncertainty in the prediction of missing data and creates multiple complete datasets. However,  MI requires conditions such as congeniality and self-efficient estimation \citep{meng1994multiple, yang2016fractional} to achieve valid estimation. As an alternative effective imputation tool, fractional imputation was proposed by \cite{kalton1984some}, and investigated by \cite{kim2004fractional} in a way of achieving efficient hot deck imputation. \cite{kim2011parametric} proposed parametric fractional imputation, which is based on parametric model assumption and is sensitive to failure of the model assumption. \cite{yang2016fractional} provide a comprehensive overview of fractional imputation. 

Nonparametric imputation, such as Kernel regression imputation, is fully robust but is not applicable when the dimension of the covariates is large due to the curse of dimensionality. Semiparametric imputation is another robust imputation method that is based on a flexible model where the number of parameters can increase with the sample size. \cite{murray2016multiple} proposed a Bayesian joint model for multiple imputation of missing values and \cite{sang2018semiparametric} developed semiparametric fractional imputation. Both methods assume Gaussian mixture models (GMM) jointly for multivariate continuous variables. 

In this paper, we propose another semiparametric imputation using a more flexible model assumption than the GMM. In the proposed mixture model, we still assume a Gaussian model for the conditional distribution of the study variable given the auxiliary variable, but the marginal distribution of the auxiliary variable is not necessarily Gaussian. Thus, our proposed imputation is more flexible than the imputation method based on GMM. For example, as demonstrated in the simulation study in Section 6, our proposed method provides more accurate prediction than the GMM under a skewed population. Thus, the resulting imputation estimator achieves smaller mean squared errors than other competitors. The computation is based on EM algorithm and it is relatively simple and fast. Furthermore, the proposed model can handle high dimensional covariates problem by incorporating penalized regression in the M-step of the EM algorithm. 

Our paper is organized as follows. After illustrating a basic setup of the problem  with a short review of some existing imputation models in Section \ref{sec:setup}, we introduce the proposed adaptive mixture models in Section \ref{sec:metd}. In Section \ref{sec:prop}, we show that the proposed model achieves a lower approximation error bound to any unknown target density based on the Kullback-Leibler divergence than the GMM. Also, we present an application of the proposed method to high-dimensional data by using the penalized maximum likelihood method in Section \ref{sec:ext}. In Section \ref{sec:simul}, two extensive simulation studies are presented to investigate the finite sample performance of the proposed imputation method. In Section \ref{sec:app}, the proposed method is applied to handle the real data problem with the 2017 Korean Household Income and Expenditure Survey (KHIES) conducted by Statistics Korea. Some concluding remarks are made in Section \ref{sec:concluding}. 

\section{Basic Setup}
\label{sec:setup}

Suppose that $\bm{x}$ and $\bm{y}$ are observed in the sample, where $\bm{y}=(y_1, \ldots, y_p)'$ is a $p$-dimensional vector of study variables and $\bm{x}=(x_1, \ldots, x_q)'$ is a $q$-dimensional vector of auxiliary variables. We assume that $\bm{y}$ is subject to missingness and $\bm{x}$ is always observed. 

Let $\bm{y}_{obs}$ and $\bm{y}_{mis}$ denote the observed and missing part of $\bm{y}$, respectively. That is, $\bm{y} = (\bm{y}_{obs}', \bm{y}_{mis}')'$. We assume the missing mechanism is missing at random in the sense of \cite{rubin1976inference}, which can be described as $f(\bm{\delta} \mid \bm{x}, \bm{y}) = f(\bm{\delta} \mid \bm{x}, \bm{y}_{obs})$, where $\bm{\delta}=(\delta_1, \ldots, \delta_p)'$ is the response indicator vector for $\bm{y}$ defined as $\delta_j =1$ if $y_j$ is observed, otherwise $\delta_j=0$. Imputation model is then the prediction model for $\bm{y}_{mis}$ and can be constructed from the conditional distribution of $\bm{y}_{mis}$ given $\bm{x}$ and $\bm{y}_{obs}$,
\begin{eqnarray}
f(\bm{y}_{mis} \mid \bm{x}, \bm{y}_{obs}) = \frac{f(\bm{y} \mid \bm{x})}{\int f(\bm{y} \mid \bm{x}) d \bm{y}_{mis}}, \label{imp.m1}
\end{eqnarray}
where we need a model assumption for $f(\bm{y} \mid \bm{x})$.

In fractional hot deck imputation \citep{kim2004fractional}, for example, the conditional distribution of $\bm{y}$ given $\bm{x}$ can be written as 
\begin{eqnarray}
f(\bm{y} \mid \bm{x}) = \sum_{g=1}^G P(z = g \mid \bm{x}) f(\bm{y} \mid z = g),
\end{eqnarray}
where $z \in \{1, \ldots, G\}$ is a cell indicator variable for imputation cells. The sample is partitioned into $G$ imputation cells so that the conditional distribution of $\bm{y}$ within the cells are homogeneous and imputed values are taken from the respondents within the same cell.

If $f(\bm{y} \mid \bm{x})$ is a parametric model with parameter $\bm{\theta}$, then the imputation can be performed in two steps: (1) estimate parameter $\bm{\theta}$, (2) perform imputation from the imputation model (\ref{imp.m1}) evaluated at the estimated parameter denoted by $\hat{\bm{\theta}}$. Parametric fractional imputation of \cite{kim2011parametric} is one example of such a procedure.

As an extension of the fractional hot deck imputation and parametric fractional imputation, \cite{sang2018semiparametric} proposed a semiparametric imputation by using multivariate Gaussian mixture models (GMM), which can be written as 
\begin{eqnarray}
f(\bm{x},\bm{y}) = \sum_{g=1}^G p_g \phi(\bm{x},\bm{y}; \bm{\mu}_g, \bm{\Sigma}_g), \label{GMM}
\end{eqnarray} 
where $0<p_1 < p_2 < \cdots < p_G < 1$ are the mixture proportions such that $\sum_{g=1}^G p_g= 1$, and $\phi(\cdot ;\bm{\mu}_g, \bm{\Sigma}_g)$ is the density of multivariate normal distribution with parameter $(\bm{\mu}_g, \bm{\Sigma}_g)$. Under this model, the conditional distribution of $\bm{y}$ given $\bm{x}$ is 
\begin{eqnarray*}
	f(\bm{y} \mid \bm{x}) = \sum_{g=1}^G P(z=g \mid \bm{x}) \phi(\bm{y} \mid \bm{x}, z=g), 
\end{eqnarray*}
where 
\begin{eqnarray*}
	P(z=g \mid \bm{x}) = \frac{p_g\phi(\bm{x} \mid z=g) }{\sum_{g=1}^G p_g \phi(\bm{x} \mid z=g)},
\end{eqnarray*}
and the conditional distribution $\phi(\bm{y} \mid \bm{x}, z=g)$ can be easily derived from the joint normality of $(\bm{x},\bm{y})$ given $z=g$. 

The GMM provides a flexible modeling, but it becomes very unstable when the dimension of $\bm{x}$ is large. Also, departure from normality introduces additional mixture components, which often lead to overfitting the model and inaccurate prediction.

\section{Proposed method}
\label{sec:metd}

We now discuss the proposed method that relaxes the assumption in (\ref{GMM}). Under complete response, we assume that 
\begin{equation} 
f( \bm{y} \mid \bm{x}) = \sum_{g=1}^G \pi_g (\bm{x}) f_g ( \bm{y} \mid \bm{x}), 
\label{CGMM}
\end{equation} 
where $ \pi_g ( \bm{x} ) = P( z=g \mid \bm{x}) $ and $f_g (\bm{y} \mid \bm{x})$ is a Gaussian distribution given $\bm{x}$ and $z=g$. 
We further assume that $\pi_g(\bm{x})= \pi_g ( \bm{x}; \bm{\alpha}) $ follows a multinomial logit model, 
\begin{equation} 
\pi_g(\bm{x}; \bm{\alpha} ) = \frac{\exp (\alpha_{g0} + \bm{x}'\bm{\alpha}_{g1}) }{\sum_{h=1}^G \exp(\alpha_{h0} + \bm{x}'\bm{\alpha}_{h1})},
\label{DRM}
\end{equation} 
with the parameter $\bm{\alpha} = \{ \bm{\alpha}_g = (\alpha_{g0}, \bm{\alpha}_{g1}')': g=1, \cdots, G, ~\alpha_{10}=0, ~\bm{\alpha}_{11} = \bm{0}_q\}$, where $\bm{0}_q$ is a $q$-dimensional zero vector. Model (\ref{CGMM}) can be called the conditional Gaussian mixture model (CGMM) and we still assume a Gaussian model for the conditional distribution $f( \bm{y} \mid \bm{x}, z=g)$.  

In fact, model (\ref{CGMM}) can be derived from the following joint model,
\begin{equation}
f(\bm{x}, \bm{y}) = \sum_{g=1}^G p_g f_1 (\bm{x} \mid z=g) f_2 ( \bm{y} \mid \bm{x}, z=g), 
\label{JointCGMM}
\end{equation} 
where $f_1$ follows the density ratio model (DRM) given by 
\begin{equation}
\log \left\{ \frac{ f_1 ( \bm{x}\mid z=g) }{ f_1 (\bm{x} \mid z=1)}  \right\} = \gamma_{g0} + \bm{x}'\bm{\gamma}_{g1},
\label{DRM1}
\end{equation} 
where $\gamma_{10}=0$ and $\bm{\gamma}_{11} = \bm{0}_q$. Under DRM in (\ref{DRM1}), the marginal distribution of $\bm{x}$ given $z=g$ is an exponential tilting of the density of $\bm{x}$ given $z=1$. The marginal density of $\bm{x}$ given $z=1$ is completely unspecified. \cite{qin1998inferences} used an empirical likelihood approach to estimate parameters under DRM. Since 
$$  \frac{ f_1 ( \bm{x}\mid z=g) }{ f_1 (\bm{x} \mid z=1)  } =  \frac{ p_1}{ p_g} \times  \frac{ P ( z=g \mid \bm{x}) }{ P(  z=1 \mid \bm{x})  }, $$
we can obtain $\alpha_{g0} = \gamma_{g0} + \log ( p_g/p_1)$ and $\bm{\alpha}_{g1} = \bm{\gamma}_{g1}$ in (\ref{DRM}). Thus, the CGMM in (\ref{CGMM}) with the multinomial logistic model (\ref{DRM}) can be derived from (\ref{JointCGMM}) with DRM assumption in (\ref{DRM1}). The DRM assumption in (\ref{DRM1}) covers a broader class of distributions that includes the Gaussian distribution as a special case. Therefore, the proposed method is more flexible than the GMM method. 

For parameter estimation under complete response, we can use the following EM algorithm. 
\begin{enumerate}
	\item{[E-step]} Given the current parameter values, compute 
	\begin{eqnarray*}
		\pi_{ig}^{(t) } &=& P ( z_i = g \mid \bm{x}_i, \bm{y}_i; \bm{\theta}^{(t)} ) \\
		&=& \frac{\pi_g(\bm{x}_i; \bm{\alpha}^{(t)} ) f_2 ( \bm{y}_i \mid \bm{x}_i, z_i=g ; \bm{\psi}_g^{(t)} )} { \sum_{g=1}^G  \pi_g(\bm{x}_i; \bm{\alpha}^{(t)} ) f_2 ( \bm{y}_i \mid \bm{x}_i, z_i = g  ; \bm{\psi}_g^{(t)} )} 
	\end{eqnarray*} 
	where $\bm{\psi}_g$ is the parameter in the conditional distribution $ f_2 ( \bm{y}_i \mid \bm{x}_i, z_i=g )$ and $\bm{\theta}$ is all the parameters, $\bm{\theta} = \{ \bm{\alpha}_g, \bm{\psi}_g: g=1, \ldots, G; \bm{\alpha}_1 =\bm{0} \}$.

	\item{[M-step]} Using $\pi_{ig}^{(t)}$, update the parameters by solving 
	$$  \sum_{i=1}^n \left\{ \pi_{ig}^{(t)} - \pi_g ( \bm{x}_i; \bm{\alpha}_g)  \right\} (1, \bm{x}_i')'= \bm{0}$$
	and 
	$$ \sum_{i=1}^n \pi_{ig}^{(t)}\left\{  \frac{\partial}{ \partial \bm{\psi}_g }  \log f_2( \bm{y}_i \mid \bm{x}_i, z_i = g; \bm{\psi}_g)\right\}  = \bm{0}. $$
\end{enumerate}

Under the existence of missing data, the imputation model under CGMM in (\ref{CGMM}) is 
\begin{equation}
f( \bm{y}_{mis} \mid \bm{x}, \bm{y}_{obs}) = \sum_{g=1}^G \pi_g (\bm{x}, \bm{y}_{obs}) f_g ( \bm{y}_{mis} \mid \bm{x}, \bm{y}_{obs}),
\label{9}
\end{equation} 
where 
$$ \pi_g (\bm{x}, \bm{y}_{obs}) = \frac{\pi_g (\bm{x}) f_2 ( \bm{y}_{obs} \mid \bm{x}, z=g) }{ \sum_{g=1}^G\pi_g (\bm{x}) f_2 ( \bm{y}_{obs} \mid \bm{x}, z=g)}
$$
and
$$
f_g(\bm{y}_{mis} \mid \bm{x}, \bm{y}_{obs}) = \frac{f_2(\bm{y} \mid \bm{x}, z=g)}{\int f_2(\bm{y} \mid \bm{x}, z=g) d \bm{y}_{mis}}.
$$
Note that $ f_2 ( \bm{y}_{obs} \mid \bm{x}, z=g)= \int f_2 ( \bm{y} \mid \bm{x}, z=g) d\bm{y}_{mis}$ is still a Gaussian distribution. 
The EM algorithm under missing data can be described as follows:

\begin{enumerate}
	\item{[E-step]} Given the current parameter values, compute
	\begin{eqnarray*}
		\pi_{ig}^{(t) } &=& P ( z_i = g \mid \bm{x}_i, \bm{y}_{i, obs}; \bm{\theta}^{(t)} ) \\
		&=& \frac{\pi_g(\bm{x}_i; \bm{\alpha}^{(t)} ) f_2 ( \bm{y}_{i, obs}  \mid \bm{x}_i, z_i=g ; \bm{\psi}_g^{(t)} )} { \sum_{g=1}^G  \pi_g(\bm{x}_i; \bm{\alpha}^{(t)} ) f_2 ( \bm{y}_{i, obs}  \mid \bm{x}_i, z_i = g  ; \bm{\psi}_g^{(t)} )}
	\end{eqnarray*}
	where $\bm{\psi}_g$ is the parameter in the conditional distribution $ f_2 ( \bm{y}_{i, obs}  \mid \bm{x}_i, z_i=g )$, which is Gaussian.
	
	\item{[M-step]} Using $\pi_{ig}^{(t)}$, update the parameters by solving
	$$  \sum_{i=1}^n \left\{ \pi_{ig}^{(t)} - \pi_g ( \bm{x}_i; \bm{\alpha})  \right\} (1, \bm{x}_i' )'= \bm{0} $$
	and
	$$ \sum_{i=1}^n \pi_{ig}^{(t)}\left\{  \frac{\partial}{ \partial \bm{\psi}_g }  \log f_2( \bm{y}_{i, obs} \mid \bm{x}_i, z_i = g; \bm{\psi}_g)\right\}  = \bm{0}. $$
\end{enumerate}
We repeat this procedure until a convergence criterion meets. 

Let $\hat{\bm{\theta}} = \{\hat{\bm{\alpha}}_g, \hat{\bm{\psi}}_g: g=1, \ldots, G; \hat{\bm{\alpha}}_1=\bm{0} \}$ denote the maximum likelihood estimates obtained from the above EM algorithm. For nonresponse $\bm{y}_{i,mis}$, we compute the imputed value, denoted by $\hat{\bm{y}}_{i,mis}$, as 
\begin{eqnarray}
\hat{\bm{y}}_{i,mis} = \sum_{g=1}^G \hat{\pi}_{ig} \bm{\mu}_{ig}(\hat{\bm{\psi}}_g),
\end{eqnarray}
where ${\bm{\mu}}_{ig}(\bm{\psi}_g) = E(\bm{y}_{i,mis} \mid \bm{x}_i, \bm{y}_{i,obs}; \bm{\psi}_g)$ and 
\begin{eqnarray}
\hat{\pi}_{ig} = \frac{\pi_g(\bm{x}_i; \hat{\bm{\alpha}} ) f_2 ( \bm{y}_{i, obs}  \mid \bm{x}_i, z_i=g ; \hat{\bm{\psi}}_g )} { \sum_{g=1}^G  \pi_g(\bm{x}_i; \hat{\bm{\alpha}} ) f_2 ( \bm{y}_{i, obs}  \mid \bm{x}_i, z_i = g  ; \hat{\bm{\psi}}_g)}.
\end{eqnarray}
This is a weighted sum of the $G$ conditional cell means.

{\bf Remark 1} For the choice of $G$, Bayesian Information Criterion (BIC) of \cite{schwarz1978estimating} can be used. In our context, the BIC can be written as 
\begin{eqnarray}
BIC(G) = -2 \log L_G(\hat{\bm{\theta}}_{G}) + d_G \log(n), \label{bic}
\end{eqnarray}
where 
\begin{eqnarray*}
	\log L_G(\hat{\bm{\theta}}_G) = \sum_{i=1}^n \log \left\{\sum_{g=1}^G {\pi}_g(\bm{x}_i; \hat{\bm{\alpha}}) f(\bm{y}_{i,obs} \mid \bm{x}_i; \hat{\bm{\psi}}_g) \right\} 
\end{eqnarray*}
and $\hat{\bm{\theta}}_G = \{ \hat{\bm{\alpha}}_g, \hat{\bm{\psi}}_g : g=1, \ldots, G; \hat{\bm{\alpha}}_1 = \bm{0}\}$ is the estimated parameter of the $G$-component proposed mixture model and $d_G=dim(\hat{\bm{\theta}}_G)$. The optimal $G$ is the one that minimizes the BIC in (\ref{bic}). Instead of using BIC, we may use 10-fold cross-validation, which is more computationally extensive method for model selection. 
% Model Selection for gaussian mixture models (Huang and Zhang, Statistica Sinica)

\section{Statistical properties}
\label{sec:prop}

We discuss the accuracy of density estimation using the CGMM. To quantify the accuracy of density estimation, we define approximation error to an unknown target density function, denoted by $f^\ast$, in terms of the Kullback-Leibler (KL) divergence. For any $f \in \mathcal{C}$, where $\mathcal{C}$ is a class of density functions to approximate $f^\ast$, the approximation error of $f$ to $f^\ast$ is defined to be the KL divergence between $f^\ast$ and $f$, 
\begin{eqnarray*}
	d_{KL}(f^\ast || f) =  E_{f^\ast} \left\{\log \frac{f^\ast(\bm{x})}{f(\bm{x})}\right\},
\end{eqnarray*}
where $E_{f^\ast}\{\cdot\}$ is the expectation with respect to the distribution with density $f^\ast$. 

We start with assuming that the target density function $f^\ast$ is unknown and continuous with a compact support in $R^{p+q}$. Also, we assume $f^\ast \in \mathcal{F}$, where
\begin{eqnarray}
\mathcal{F} = \{f: \int_{\chi} f(\bm{x},\bm{y}) d(\bm{x},\bm{y}) = 1, f(\bm{x},\bm{y}) \ge \eta > 0, \forall (\bm{x},\bm{y}) \in {\chi} \}, \label{target}
\end{eqnarray}
for some positive constant $\eta$ and $\chi$ is the support of $(\bm{x},\bm{y})$. It is natural to consider densities that are positive since the KL divergence is used as a discrepancy measure between two densities. \citep{zeevi1997density}

For $(\bm{x},\bm{y}) \in \chi \subset R^{p+q}$, we define two classes of $G$-component mixtures as 
\begin{eqnarray*}
	\mathcal{C}_{0,G} &=& \left\{f: f(\bm{x},\bm{y}) = \sum_{g=1}^G p_{0,g} f_{0}(\bm{x},\bm{y}; \bm{\theta}_{0,g}), \sum_{g=1}^G p_{0,g} = 1, p_{0,g} \ge 0, \bm{\theta}_{0,g} \in \Theta_0 \right\},\\
	\mathcal{C}_{1,G} &=& \left\{f: f(\bm{x},\bm{y}) = \sum_{g=1}^G p_{1,g} f_{1}(\bm{x},\bm{y};\bm{\theta}_{1,g}), \sum_{g=1}^G p_{1,g} = 1, p_{1,g} \ge 0, \bm{\theta}_{1,g} \in \Theta_1 \right\},
	%\mathcal{C}_{1,G} &=& \left\{f_1: f_1(y,x) = \sum_{g=1}^G p_{\delta_g,x}   \phi_{\eta_g}(y \mid x)h(x), (\delta_g, \eta_g) \in \tilde{\Theta} \right\},
\end{eqnarray*}
where $\Theta_j \subset R^{d_j}$ is the parameter (product) spaces and $d_j = dim(\bm{\theta}_{j,g})$, for $j=0, 1$. Here, we use 
\begin{eqnarray*}
	f_0(\bm{x},\bm{y}; \bm{\theta}_{0,g}) &=& \phi(\bm{x}; \bm{\theta}_{0,g}^x) \phi(\bm{y} \mid \bm{x}; \bm{\theta}_{0,g}^y), \\
	f_1(\bm{x},\bm{y}; \bm{\theta}_{1,g}) &=& f_1(\bm{x}; \bm{\theta}_{1,g}^x) \phi(\bm{y} \mid \bm{x}; \bm{\theta}_{1,g}^y),
\end{eqnarray*}
where $\phi(\cdot; \bm{\theta})$ is a multivariate Gaussian density with parameter $\bm{\theta}$ and $f_1(\cdot ; \bm{\theta})$ satisfies
\begin{eqnarray*}
	\log \frac{f_1(\bm{x}; \bm{\theta}_{1,g}^x)}{f_1(\bm{x}; \bm{\theta}_{1,1}^x)} = (1, \bm{x}')\bm{\gamma}_g
\end{eqnarray*}
where $\bm{\theta}_{1,g}^x = \bm{\gamma}_g$ and $\bm{\gamma}_1 = \bm{0}$. We also define two classes of the corresponding continuous convex combinations 
\begin{eqnarray*}
	\mathcal{C}_0 &=& \left\{ \bar{f}: \bar{f}(\bm{x},\bm{y}) = \int_{\Theta_0} f_0(\bm{x},\bm{y}; \bm{\theta}_0) P_0(d \bm{\theta}_0), ~~ P_0:~ \mbox{a probability measure on}~ (\Theta_0, \mathcal{F}_0) \right\},\\
	\mathcal{C}_1 &=& \left\{\bar{f}: \bar{f}(\bm{x},\bm{y}) = \int_{\Theta_1} f_1( \bm{x}, \bm{y} ; \bm{\theta}_1) P_1(d\bm{\theta}_1), ~~ P_1: ~ \mbox{a probability measure on}~ (\Theta_1, \mathcal{F}_1) \right\},
\end{eqnarray*}
where $(\Theta_j, \mathcal{F}_j)=(\Theta_j^x \times \Theta_j^y, \mathcal{F}_j^x \times \mathcal{F}_j^y)$ is the product measurable space, and
$(\Theta_j^x, \mathcal{F}_j^x, P_j^x)$ and $(\Theta_j^y, \mathcal{F}_j^y, P_j^y)$ are two parameter measure spaces for $\bm{\theta}^x$ and $\bm{\theta}^y$, respectively. That is, $P_j= P_j^x \times P_j^y$ is a product measure on $(\Theta_j, \mathcal{F}_j)$, for $j=0, 1$.

\cite{li2000mixture} derived an explicit form of the approximation error bound for a finite mixture density based on the KL divergence. Lemma 1 presents the approximation error bounds of the $G$-component mixture densities in the classes $\mathcal{C}_{0,G}$ and $\mathcal{C}_{1,G}$, respectively.

{\bf Lemma 1} Suppose that a target density function $f^\ast$ belongs to $\mathcal{F}$ in (\ref{target}). Let $f_{j,G} \in \mathcal{C}_{j,G}$ for $j \in \{0, 1\}$. For any given $G$, the approximation error of $f_{j,G}$ to $f^\ast$  is bounded from above as follows, 
\begin{eqnarray}
d_{KL}(f^\ast || f_{j,G}) \le	d_{KL}(f^\ast || \bar{f}_j) + \frac{c_{f^\ast,j}^2 \kappa_j}{G},  \label{lem1}
\end{eqnarray}
where $\bar{f}_j \in \mathcal{C}_j$ and 
\begin{eqnarray*}
	c_{f^\ast,j}^2  &=& \int \frac{\int f_j(\bm{x},\bm{y}; \bm{\theta}_j)^2 P_j(d\bm{\theta}_j)}{\{ \int f_j(\bm{x},\bm{y}; \bm{\theta}_j) P_j(d \bm{\theta}_j)\}^2} f^\ast(\bm{x},\bm{y}) d(\bm{x},\bm{y}),\\
	%\gamma_j &\propto& \underset{\theta_{j}, \tilde{\theta}_{j}, x, y}{sup} \log \frac{f_j(x,y; \theta_{j})}{f_j(x,y; \tilde{\theta}_{j})}.
	\kappa_j &\propto& \underset{\bm{\theta}_{j}, \tilde{\bm{\theta}}_{j} \in \Theta_j, (\bm{x}, \bm{y}) \in \chi} {\sup} \log \frac{f_j(\bm{x},\bm{y}; \bm{\theta}_{j})}{f_j(\bm{x},\bm{y}; \tilde{\bm{\theta}}_{j})}.
\end{eqnarray*}

Lemma 1 shows that the rate of convergence is $1/G$ and the constants in the upper bound, $c_{f^\ast,j}^2$ and $\kappa_j$, depend on the choices of $\mathcal{C}_j$ and the target density $f^\ast$. By using the approximation error bound of Lemma 1, we compare the quality of approximation to an unknown target density between the two classes $\mathcal{C}_{0,G}$ and $\mathcal{C}_{1,G}$ for any given $G$ in the following theorem.

{\bf Theorem 1} For any arbitrary target density $f^\ast \in \mathcal{F}$ and any $\epsilon>0$, 
consider $\bar{f}_0 \in \mathcal{C}_0$ and $\bar{f}_1 \in \mathcal{C}_1$ satisfying $d_{KL}(f^\ast || \bar{f}_0) = d_{KL}(f^\ast || \bar{f}_1)=\epsilon$. Then, it holds that for any given $G$,
\begin{eqnarray*}
	b_{f^\ast}(f_{0,G}) \ge b_{f^\ast} (f_{1,G}),
\end{eqnarray*}
where $b_{f^\ast}(f_{j,G}) = d_{KL}(f^\ast || \bar{f}_j) + c_{f^\ast,j}^2 \kappa_j / G$ is the upper bound of $d_{KL}(f^\ast || f_{j,G})$ obtained from Lemma 1.

The proof of Theorem 1 is presented in the supplementary material (Section S1). Theorem 1 shows that for any target density function $f^\ast \in \mathcal{F}$, the proposed mixture densities using CGMM achieve a lower approximation error bound than the Gaussian mixtures under the same number of components.

We now investigate the approximate error bound for the maximum likelihood estimator of the proposed mixture density, $f_{1,G} \in \mathcal{C}_{1,G}$ under complete response. Denote the maximum likelihood estimator by $\hat{\bm{\theta}}$ defined as $\hat{\bm{\theta}} = \mathrm{argmax}_{\bm{\theta}} \ell_{n}(\bm{\theta})$,  
where 
\begin{eqnarray*}
	\ell_{n}(\bm{\theta}) = \frac{1}{n}\sum_{i=1}^n \log f_{1,G}(\bm{x}_i,\bm{y}_i; \bm{\theta}),
\end{eqnarray*}
and $f_{1,G}(\bm{x},\bm{y}; \bm{\theta}) = \sum_{g=1}^G \pi_{1,g} f_{1}(\bm{x},\bm{y}; \bm{\theta}_{1,g})$. Let $\hat{f}_{1,G}$ denote the value of $f_{1,G}(\bm{x}, \bm{y}; \bm{\theta})$ evaluated at $\bm{\theta} = \hat{\bm{\theta}}$.

{\bf Theorem 2} Under the assumptions (A1) - (A7) stated in the supplementary material (Section S2), it holds that for any $\epsilon_1 >0$ and $G$, 
\begin{eqnarray*}
	E_{f_n^\ast}\{ d_{KL}(f^\ast || \hat{f}_{1,G})\}= \epsilon_1 + \frac{c_{f^\ast, 1}^2 \kappa_1}{G} + O \left(\frac{m}{n}\right),
\end{eqnarray*}
for sufficiently large $n$, where $m=Tr(H_G(\bm{\theta}^0)^{-1}J_G{(\bm{\theta}^0)})$, $\bm{\theta}^0$ is the maximizer of $E_{f^\ast}\{\log f_{1,G}(\bm{x},\bm{y}; \bm{\theta}) \}$, and 
\begin{eqnarray*}
	H_G(\bm{\theta}) &=& - E_{f^\ast} \left\{\frac{\partial^2 \log f_{1,G}(\bm{x}, \bm{y} ; \bm{\theta})}{\partial \bm{\theta} \partial \bm{\theta}'}  \right\},\\
	J_G(\bm{\theta}) &=& Var_{f^\ast} \left\{ \frac{\partial \log f_{1,G}(\bm{x}, \bm{y} ; \bm{\theta})}{\partial \bm{\theta} } \right\}.
\end{eqnarray*}

Theorem 2 implies that there exists $G$ such that for any $\epsilon > 0$, 
\begin{eqnarray*}
	\|f^\ast - \hat{f}_{1,G}\|_1 < \epsilon ~~~~ a.e. (\mu_{f_n^\ast}),
\end{eqnarray*}
for sufficiently large $n$, where $\|f^\ast - \hat{f}_{1,G}\|_1 = \int | \hat{f}_{1,G} (\bm{x}, \bm{y})  - f^\ast (\bm{x},\bm{y})| d \lambda(\bm{x},\bm{y})$, and $\mu_{f_n^\ast}$ is a probability measure generated by the true probability density function of $\hat{\bm{\theta}}$, denoted by $f_n^\ast$. See Section S2 in the supplementary material for the proof.

{\bf Remark 2} For sufficiently large $G$, $f_{1,G}^0$ converges to the true density function to the $\epsilon_1$-specified accuracy by (S.6) in the supplementary material (Section S2), and $H_G(\bm{\theta}^0)^{-1} J_G(\bm{\theta}^0) \approx I_{Gd_1 \times Gd_1}$, where $d_1=dim(\bm{\theta}^0)$. Therefore, 
\begin{eqnarray}
E_{f^\ast_n} \{ d_{KL}(f^\ast || \hat{f}_{1,G}) \} = \epsilon_1 + \frac{c_{f^\ast, 1}^2 \kappa_1}{G} + O \left(\frac{G d_1}{n}\right), \label{est.errb}
\end{eqnarray}
for sufficiently large $n$ and $G$. A similar argument is used in \cite{zeevi1997density}. 

{\bf Remark 3} Theorem 2 and Remark 2 also hold for the Gaussian mixture density, $f_{0,G} \in \mathcal{C}_{0,G}$. By using (\ref{est.errb}), Theorem 1 and the fact of $d_0 > d_1$, where $d_0$ denotes the dimension of the parameters specified in $f_{0,G}$, we can show that the proposed mixture density achieves a lower approximation error bound than the GMM in terms of the KL divergence measure.

We finally establish some asymptotic behaviors of the imputed estimator based on the proposed mixture model under item nonresponse. Suppose that we are interested in estimating a target parameter, denoted by $\bm{\xi}=(\xi_1, \ldots, \xi_K)'$, defined as the solution to $E_{f^\ast}\{U(\bm{\xi}; \bm{x}, \bm{y})\} = \bm{0}$. Without item nonresponse, a consistent estimator of $\bm{\xi}$ is obtained by solving 
\begin{eqnarray*}
	\frac{1}{n}\sum_{i=1}^n U(\bm{\xi}; \bm{x}_i, \bm{y}_i) = \bm{0}. 
\end{eqnarray*}
Under missing data, our proposed estimator of $\bm{\xi}$ is computed by solving 
\begin{eqnarray*}
	\frac{1}{n}\sum_{i=1}^n E_{\hat{f}_{1,G}} \{ U(\bm{\xi}; \bm{x}_i, \bm{y}_i)  \mid \bm{x}_i, \bm{y}_{i,obs} \} = \bm{0}, 
\end{eqnarray*}
where $E_{\hat{f}_{1,G}} \{ U(\bm{\xi}; \bm{x}, \bm{y}) \mid \bm{x}, \bm{y}_{obs} \} $ is the conditional expectation with respect to $\hat{f}_{1,G}(\bm{y}_{mis} \mid \bm{x}, \bm{y}_{obs};\hat{\bm{\theta}})$. Here, $\hat{\bm{\theta}} = \mathrm{argmax}_{\bm{\theta}} \ell_{obs}(\bm{\theta})$, and 
\begin{eqnarray*}
	\ell_{obs}(\bm{\theta}) = \sum_{i=1}^n \log \left\{\sum_{g=1}^G {\pi}_g(\bm{x}_i; \bm{\alpha}) f(\bm{y}_{i,obs} \mid \bm{x}_i; \bm{\psi}_g) \right\}.
\end{eqnarray*}

{\bf Theorem 3} For $f^\ast, f_{1,G} \in \mathcal{F}$, let $G = \epsilon^{-\tau}$ such that $\| \hat{f}_{1,G} - f^\ast \|_1 < \epsilon ~~ a.e.(\mu_{f_n^\ast})$ for any small $\epsilon > 0$ and $\tau > 0$. Under the assumptions stated in the supplementary material (Section S3),
\begin{eqnarray*}
	\sqrt{n}(\hat{\bm{\xi}} - \bm{\xi}^0) \rightarrow N(\bm{0}, V),
\end{eqnarray*} 
where $V$ is positive definite, and $\bm{\xi}^0$ satisfies $E_{f^\ast} \{U(\bm{\xi} ; \bm{x}, \bm{y}) \}  = \bm{0}$. See Section S3 in the supplementary material for the proof.

\section{Extension}
\label{sec:ext}

In many practical situation, the dimension of $y$ can be small but the dimension of covariates $\bm{x}$ can be large. In this case, the imputation using GMM can have numerical problems and the prediction can be unstable. Under our CGMM setup, we can use a penalized regression method to select some important covariates so that the prediction accuracy can get improved.  

For simplicity, assume that $y \in R$ and $\bm{x} \in R^q$. We define a penalized log-likelihood function with full observation $\{(\bm{x}_i, y_i, z_i): i=1, \ldots, n\}$ as 
\begin{eqnarray*}
	\log L_p(\theta \mid \bm{x}, y, z) &=& \sum_{i=1}^n \sum_{g=1}^G I(z_i=g) \{\log Pr(z_i=g \mid \bm{x}_i; \bm{\alpha}) + \log f(y_i \mid \bm{x}_i, z_i=g; \bm{\beta}_g, \sigma_g^2 ) \} \\
	&~& - \sum_{g=1}^G P_{\lambda}(\bm{\alpha}_g, \bm{\beta}_g),
\end{eqnarray*} 
where $\bm{\alpha}_1=\bm{0}$ and $P_{\lambda}(\bm{\alpha}_g, \bm{\beta}_g)$ is a penalty function on $\bm{\alpha}_g$ and $\bm{\beta}_g$ such as the LASSO \citep{tibshirani1996regression}, ridge, mixture of the two called the elastic net \citep{zou2005regularization}, SCAD \citep{fan2001variable} and so on. In this study, we apply the lasso ($L_1-$ norm) penalty given by 
\begin{eqnarray*}
	P_{\lambda}(\bm{\alpha}_g, \bm{\beta}_g) = \lambda \sum_{j=1}^{q} ( | \alpha_{g,j} | + |\beta_{g,j}|). 
\end{eqnarray*}

The corresponding expected log-likelihood function, denoted by ${\ell}_p(\bm{\theta})$, is 
\begin{eqnarray*}
	{\ell}_p(\bm{\theta}) &=& E\{ \log L_p(\bm{\theta} \mid \bm{x}, y, z) \mid \bm{x}, y \} \notag \\
	&=& \sum_{i=1}^n  \sum_{g=1}^G \pi_{ig} \{ \log Pr(z_i = g \mid \bm{x}_i; \bm{\alpha}) + \log f(y_i \mid \bm{x}_i, z_i=g;  \bm{\beta}_g, \sigma_g^2) \} \\
	&~& - \sum_{g=1}^G P_{\lambda}(\bm{\alpha}_g, \bm{\beta}_g)
\end{eqnarray*}
where 
\begin{eqnarray}
\pi_{ig} = Pr(z_i = g \mid \bm{x}_i, y_i) = \frac{Pr(z_i = g \mid \bm{x}_i) f(y_i \mid \bm{x}_i, z_i=g)}{\sum_{g=1}^G Pr(z_i = g \mid \bm{x}_i)f(y_i \mid \bm{x}_i, z_i=g) }. \label{ch5.pig}
\end{eqnarray}

We can use the penalized maximization in the M-step of the EM algorithm. That is, the E-step remains the same. In the M-step, we update $\bm{\beta}$ by maximizing $\ell_p^{(t)}(\bm{\beta})$, where 
\begin{eqnarray}
\ell_p^{(t)}(\bm{\beta}) =  -\frac{1}{2 \sigma_g^{2(t)}} \sum_{i=1}^n \sum_{g=1}^G \pi_{ig}^{(t)} \{ y_i - (1,\bm{x}')\bm{\beta}_g\}^2 - \lambda \sum_{g=1}^G \sum_{j=1}^q |\beta_{g,j}|, \label{obj1}
\end{eqnarray}
and $\pi_{ig}^{(t)}$ is obtained from (\ref{ch5.pig}) using the current parameter values. 

To find the maximizer of (\ref{obj1}), we can use the cyclic coordinate descent algorithm described in \cite{friedman2010regularization}. Suppose that we update $\beta_{g,\ell}^{(t+1)}$ for $\ell \neq j$ and $g$. We partially optimize (\ref{obj1}) with respect to $\beta_{g,j}$. If $\beta_{g,j}^{(t+1)} > 0$, the gradient at $\beta_{g,j} = \beta_{g,j}^{(t+1)}$ is 
\begin{eqnarray*}
	\frac{\partial \ell_p^{(t)}(\beta)}{ \partial \beta_{g,j}} = \frac{1}{ \sigma_g^{2(t)}} \sum_{i=1}^n \pi_{ig}^{(t)} (y_i - (1, \bm{x}')\bm{\beta}^{(t+1)}) x_{ij} + \lambda, 
\end{eqnarray*}
and a similar expression exists if $\beta_{g,j}^{(t+1)} < 0$. Then, the coordinate-wise update for $\beta_{g,j}$ can be computed as follows: for $j=1, \ldots, q$, 
\begin{eqnarray*}
	\beta^{(t+1)}_{g,j} = \frac{S\left(\sum_{i=1}^n \pi_{ig}^{(t)} (y_i - \tilde{y}^{(t+1)}_{ig,j}) x_{ij} , \lambda\right)}{\sum_{i=1}^n  \pi_{ig}^{(t)} x_{ij}^2},
\end{eqnarray*}
where $\tilde{y}_{ig,j}^{(t+1)} = \beta_{g,0}^{(t+1)} + \sum_{\ell \neq j} x_{i\ell} \beta_{g,\ell}^{(t+1)}$ is the fitted value excluding the contribution from $x_{ij}$ and $S(z, \gamma)$ is  the soft-thresholding operator with value; $S(z, \gamma) = z-\gamma$ {if} $z > 0$ {and} $\gamma < |z|$, $z+\gamma$ if $z < 0$ and $\gamma < |z|$, otherwise 0. 
%\begin{eqnarray*}
%	S(z, \gamma) = \begin{cases} z-\gamma ~~ \mbox{if} ~~ z > 0 ~~ \mbox{and} ~~ \gamma < |z|, \\ z+\gamma ~~ \mbox{if} ~~ z < 0 ~~ \mbox{and} ~~ \gamma  < |z|,\\
%		0 ~~~~~~~~ \mbox{if} ~~ \gamma \ge |z|.
%	\end{cases}
%\end{eqnarray*}

Similarly, we update $\bm{\alpha}$ by maximizing 
\begin{eqnarray*}
	\ell_p^{(t)}(\bm{\alpha}) = {\ell}^{(t)}(\bm{\alpha})  - \lambda \sum_{g=2}^G \sum_{j=1}^q |\alpha_{g,j}| ,
\end{eqnarray*}
with respect to $\bm{\alpha}_g$ for $g=2, \ldots, G$, where
\begin{eqnarray}
{\ell}^{(t)}(\bm{\alpha}) = \sum_{i=1}^n \sum_{g=2}^G \pi_{ig}^{(t)} ((1,\bm{x}_i') \bm{\alpha}_g) - \log \left\{1+ \sum_{g=2}^G {\exp}( (1,\bm{x}_i') \bm{\alpha}_g ) \right\}.
\end{eqnarray}

As in \cite{friedman2010regularization}, we use partial Newton steps by forming a partial quadratic approximation to ${\ell}^{(t)}(\bm{\alpha})$ at $\bm{\alpha}^{(t)}$, which is given by 
\begin{eqnarray*}
	\tilde{\ell}^{(t)}(\bm{\alpha}_g) = - \frac{1}{2}\sum_{i=1}^n  \omega_{ig}^{(t)} (h_{ig}^{(t)} - (1, \bm{x}_i')\bm{\alpha}_g)^2 + C(\bm{\alpha}^{(t)}),
\end{eqnarray*}
where 
\begin{eqnarray*}
	h_{ig}^{(t)} &=& (1, \bm{x}_i') \bm{\alpha}_g^{(t)} + \frac{ \pi_{ig}^{(t)} - {p}^{(t)}_g(\bm{x}_i)}{{p}^{(t)}_g (\bm{x}_i) (1-{p}^{(t)}_g(\bm{x}_i))},\\
	\omega_{ig}^{(t)} &=& {p}^{(t)}_g(\bm{x}_i) (1-{p}^{(t)}_g(\bm{x}_i)),\\
	{p}^{(t)}_g (\bm{x}_i) &=& \frac{\exp((1,\bm{x}_i')\bm{\alpha}^{(t)}_g)} {1+\exp((1,\bm{x}_i')\bm{\alpha}^{(t)}_g)},
\end{eqnarray*}
and $C({\alpha^{(t)}})$ is a constant in terms of $\alpha_g$, for each $g$. We find a maximizer of the partial quadratic approximation, denoted by $\tilde{\ell}_p^{(t)}(\alpha_g)$, where
\begin{eqnarray*}
	\tilde{\ell}^{(t)}_p(\bm{\alpha}_g) = \tilde{\ell}^{(t)}(\bm{\alpha}_g) + \lambda  \sum_{j=1}^q |\alpha_{g,j}|,
\end{eqnarray*}
by using the coordinate descent algorithm. The coordinate-wise update for $\alpha_{g,j}$ is computed as
\begin{eqnarray*}
	\alpha^{(t+1)}_{g,j} = \frac{S\left(\sum_{i=1}^n  \omega_{ig}^{(t)} x_{ij} (h_{ig}^{(t)} - \tilde{h}^{(t+1)}_{ig, j}), \lambda\right)}{\sum_{i=1}^n  \omega_{ig}^{(t)} x_{ij}^2 },
\end{eqnarray*}
where $\tilde{h}^{(t+1)}_{ig,j} = {\alpha}^{(t+1)}_{g,0} + \sum_{\ell \neq j} {x}_{i\ell} {\alpha}^{(t+1)}_{g,\ell}$. 

We choose the tuning parameter among some  possible values, for example, roughly between 0.1 and 100 in our simulation study in Section 6, through the 10-fold cross-validation. See Section S4 in the supplementary material for the computational detail under the existence of missing data. 

\section{Simulation Study}
\label{sec:simul}
We conduct two simulation studies to evaluate the performance of the proposed method and to compare with the semiparametric imputation using Gaussian mixture models under two scenarios: (i) when a small number of covariates are given; (ii) when a relatively large number of covariates are given. 

\subsection{Simulation One}
\label{sub:simul1}
We consider four data generating models given below. 

\begin{itemize}
	\item [(i)] Model 1 (GMM $(\bm{x},y)$): We generate $\bm{x}=(x_1, x_2)$ and $y$ from a Gaussian mixture model as follows. For $g=1, 2, 3$,	
	\begin{eqnarray*}
		P(z=g) &=& \lambda_g\\
		(x_1, x_2, y)' \mid z=g & \sim & N(\bm{\mu}_g, \bm{\Sigma}),
	\end{eqnarray*}
	where we set $(\lambda_1, \lambda_2, \lambda_3) = (0.4, 0.3, 0.3)$, and $\mu_1 = (0,-2,1)', \mu_2 = (2,0,3)'$ and $\mu_3 = (-2, 2, -3)'$. Also, we set $\bm{\Sigma}_{(i,j)} = (-0.2)^{|i-j|}$, where $\bm{\Sigma}_{(i,j)}$ is the $(i,j)$th element of $\bm{\Sigma}$. 
	
	\item [(ii)] Model 2 (GMM $\bm{x}$, GMM $y$ given $\bm{x}$): We generate $\bm{x}$ from a Gaussian mixture model with $4$ components and generate $y$ from a conditional Gaussian mixture model with $2$ components given $\bm{x}$, as follows. For $g=1, \ldots, 4$, and $h=1, 2$,
	\begin{eqnarray*}
		P(z=g) &=& \lambda_g,\\
		({x}_{1}, {x}_{2})' \mid z=g &\sim& N\left(\mu_g,  \begin{bmatrix} 
			1 & 0.1 \\
			0.1 & 1
		\end{bmatrix}\right),\\
		U &=& \alpha_0 + \alpha_1 x_1 + \alpha_2 x_2 + N(0,1),\\
		y \mid U \in I_h &\sim & N((1,\bm{x}')\bm{\beta}_h, 1),
	\end{eqnarray*}
	where $I_1 = (-\infty, c), I_2 = [c, \infty)$, and $c$ is specified as the 60$\%$ sample quantile of $U$. We set 
	$(\lambda_1, \lambda_2, \lambda_3, \lambda_4) = (0.2, 0.3, 0.2, 0.3)$, and $\bm{\mu}_1 = (-1,0.5)'$, $\bm{\mu}_2 = (1,1)'$, $\bm{\mu}_3 = (0.5, -1)'$ and $\bm{\mu}_4=(0,0)'$. Also, we set $\bm{\alpha} = (1,1,0.5)'$,  $\bm{\beta}_1 = (1,2,-2)'$ and $\bm{\beta}_2 = (-1,0.5, -0.5)'$. 
	
	\item [(iii)] Model 3 (Skewed $\bm{x}$, GMM $y$ given $\bm{x}$): We use the same model as in Model 2 except for $\bm{x}$:
	\begin{eqnarray*}
		({x}_{1}, {x}_{2})' \mid z=g &\sim& LN\left(\bm{\mu}_g,  \begin{bmatrix} 
			0.5 & 0 \\
			0 & 0.5
		\end{bmatrix}\right),
	\end{eqnarray*}
	for $g=1, \ldots, 4$.
	
	\item [(iv)] Model 4 (Skewed $\bm{x}$, Skewed $y$ given $\bm{x}$): We use the same model as in Model 3 except for $y$: for $h=1, 2$,
	\begin{eqnarray*}
		y = (1,\bm{x}')\bm{\beta}_h + e, ~~~~ \mbox{if}~~ U \in I_h,
	\end{eqnarray*}
	where $e \sim Gamma(1,1)$.
\end{itemize}

We generate 1,000 finite population data with the population size, $N=20,000$ and select a sample of size $n$ equal to 1,000 by using simple random sampling from each finite population. Once the full sample is selected, we generate $\delta_i \sim Bernoulli(q_i)$ for $i=1, \ldots, n$, where  $logit(q_{i}) = -0.5 + 0.5 x_{1i}$. We assume that $y_i$ are observed only when $\delta_i=1$. The overall missing rate is about $40\%$.  

For each realized incomplete samples, we use the following methods to impute the missing values and compare their imputation accuracy. 

\begin{enumerate}
	
	\item  (PMM) Predictive-Mean Matching : Commonly used for multiple imputation using the chained equations process \citep{buuren2010mice}. An iterative method imputing missing values using linear regression. Implemented using the MICE package in R. 
	
	\item  (GMM) { Gaussian Mixture Model}  : The number of components $G$ is selected using the BIC. We consider $G \in \{1, \ldots, 10\}$. 
	
	\item  (CGMM) {Conditional Gaussian Mixture Model} :  The number of components $G$ is selected using the BIC. We consider $G \in \{1, \ldots, 10 \}$.
\end{enumerate}

To evaluate the imputation accuracy of each method, we compute the mean absolute error (MAE) and root mean squared prediction error (RMSPE) metrics defined as follows: 
\begin{eqnarray*}
	MAE &=& \frac{1}{\sum_{i=1}^n  (1-\delta_{i})} \sum_{i=1}^n   (1-\delta_{i}) | \hat{y}_{i}^\ast -  {y}_{i}|,\\
	RMSPE &=& \sqrt{\frac{1}{\sum_{i=1}^n  (1-\delta_{i})} \sum_{i=1}^n  (1-\delta_{i}) (\hat{y}_{i}^\ast -  {y}_{i})^2}, 
\end{eqnarray*}
where $\hat{y}_{i}^\ast$ is the imputed value of missing $y_{i}$ with $\delta_{i}=0$ and $y_{i}$ is the true value. Also, to compare the estimation quality, we compute the Monte Carlo mean squared error, variance and bias of each estimator of $\theta = \bar{Y}_N$, denoted by $\hat{\theta}$, where $\bar{Y}_N = N^{-1}\sum_{i=1}^N y_i$ is the finite population mean and 
\begin{eqnarray*}
	\hat{\theta} = \frac{1}{n}\sum_{i=1}^n \{  \delta_{i} y_{i} + (1-\delta_{i}) \hat{y}_{i}^{\ast}\}.
\end{eqnarray*}

Table \ref{ch5.tb1} presents the average RMSPE and MAE of the three imputation methods across the 1,000 Monte Carlo samples for each data generating model. For Model 1, CGMM and GMM are comparable, however, for Model 2, Model 3, and Model 4, CGMM has lower values of the RMSPE and MAE. 

\begin{table}
	\caption{Average root mean squared prediction error (RMSPE) and mean absolute error (MAE) of three imputation methods based on 1,000 Monte Carlo samples}
	\label{ch5.tb1}
	\begin{center}
		\begin{tabular}{cccc}
			{True Model} & {Method} & {RMSPE} & {MAE}\\
			\hline
			& PMM & 1.6572 & 1.3252 \\ 
			{Model 1}&GMM & 1.1951 & 0.9073 \\ 
			&CGMM & 1.2056 & 0.9128 \\ 
			\hline
			& PMM & 1.6913 & 1.3426 \\ 
			{Model 2}&GMM &  {1.5650} & {1.2294}\\
			&CGMM & {1.4697} & {1.1305} \\
			\hline
			& PMM & 1.9692 & 1.5470 \\ 
			{Model 3} &GMM & 1.5244 & 1.1839 \\ 
			&CGMM & 1.4131 & 1.0623 \\ 
			\hline
			& PMM & 1.9717 & 1.5372 \\ 
			{Model 4}&GMM & 1.5228 & 1.1442 \\ 
			&CGMM & 1.4188 & 1.0024 \\ 
		\end{tabular}
	\end{center}
\end{table}

\begin{table}
	\caption{Monte Carlo bias, variance and mean squared error (MSE)  of the three imputed estimators of $\bar{Y}_N$, based on 1,000 Monte Carlo samples}
	\label{ch5.tb2.add1}
	\begin{center}
		\begin{tabular}{ccccc}
			True Model & Method & Bias$~(\times 100)$ & Var$~(\times 100)$ & MSE$~(\times 100)$ \\ 
			\hline
			\multirow{4}{*}{Model 1} & Full & -0.003 & 0.637 & 0.637 \\ 
			& PMM & 0.631 & 0.792 & 0.796 \\ 
			& GMM & 0.140 & 0.703 & 0.704 \\ 
			& CGMM  & 0.099 & 0.711 & 0.711 \\ 
			\hline
			\multirow{4}{*}{Model 2} & Full  & 0.209 & 0.506 & 0.506 \\ 
			& PMM & 0.370 & 0.702 & 0.703 \\ 
			&  GMM & 0.541 & 0.653 & 0.656 \\ 
			& CGMM & 0.339 & 0.647 & 0.648 \\ 
			\hline
			\multirow{4}{*}{Model 3} & Full & 0.088 & 0.661 & 0.662 \\ 
			& PMM & 1.597 & 0.933 & 0.958 \\ 
			& GMM & 2.584 & 0.880 & 0.946 \\ 
			& CGMM & 0.145 & 0.808 & 0.808 \\ 
			\hline
			\multirow{4}{*}{Model 4} & Full& 0.284 & 0.590 & 0.591 \\ 
			& PMM & 2.246 & 0.991 & 1.041 \\ 
			& GMM & 2.948 & 0.925 & 1.012 \\ 
			& CGMM & 0.178 & 0.792 & 0.793 \\ 
		\end{tabular}
	\end{center}
	\vspace{0.05cm}
	NOTE: ``Full" indicates the full sample estimation when missing values do not exist. 
\end{table}

\begin{table}
	\caption{Observed coverage rates of the proposed imputation estimator (CGMM) for $95\%$ confidence intervals}
	\label{ch5.tb2.add2}
	\begin{center}
		\begin{tabular}{ccccc}
			True Model & Model 1 & Model 2 & Model 3 & Model 4\\ 
			\hline
			Coverage rate ($\%$) & $95.7$ & $95.4$ & $94.6$ & $95.8$\\
		\end{tabular}
	\end{center}
\end{table}

Table \ref{ch5.tb2.add1} shows the Monte Carlo mean squared errors (MSE), variances and biases of the three imputed  estimators of $\bar{Y}_N$. For all the data generating models, the imputed estimator using CGMM has lower MSE than the two competitors. Especially, for skewed distributed data such as Model 3 and Model 4, the imputed estimators using PMM and GMM show non-negligible biases, however, the imputed estimator using CGMM is almost unbiased. In addition to point estimation, confidence intervals are computed using the jackknife variance estimation. Table \ref{ch5.tb2.add2} presents the coverage rates of confidence intervals which are computed using normal approximation. It shows that the coverage rates are close to the nominal coverage level. 

\subsection{Simulation Two}
\label{sub:simul2}
We repeat the same simulation study as in Simulation One but allow for data to be generated with a higher dimension. We consider the following models called Model 5 and Model 6 to generate $\bm{x} = (x_{1}, \ldots, x_{q})'$ and $y$, where we set $q = 15$.  

\begin{itemize}
	\item [(i)] Model 5 (GMM $\bm{x}$, GMM $y$ given $\bm{x}$) : For $g=1, \ldots, 4$, and $h=1, 2$,
	\begin{eqnarray*}
		P(z=g) &=& \lambda_g,\\
		\bm{x} \mid z=g &\sim& N\left(\bm{\mu}_g, \bm{\Sigma}\right),\\
		U &=& (1, \bm{x}') \bm{\alpha} + N(0,1),\\
		y \mid U \in I_h &\sim & N((1,\bm{x}')\bm{\beta}_h, 1), ~~~ h =1, 2,
	\end{eqnarray*}
	where $I_1 = (-\infty, c), I_2 = [c, \infty)$, and $c$ is specified as the 60$\%$ sample quantile of $U$. We set
	$(\lambda_1, \lambda_2, \lambda_3, \lambda_4) = (0.2,0.3,0.2,0.3)$, $(\mu_1, \mu_2, \mu_3, \mu_4)'=(1, 2, -1, -2) \bm{1}_{q}$, and $\bm{\Sigma}_{(i,j)} = 0.5^{|i-j|}$, where $\bm{1}_q$ denotes the $q$-dimensional one vector and $\bm{\Sigma}_{(i,j)}$ is the $(i,j)$th element of $\bm{\Sigma}$. Also, we specify $\bm{\alpha}=(1,1,0,1,0,1, \bm{0}_{q-5})'$ and $\bm{\beta}_1 = (-1, 0, 2.5, 0, 3, \bm{0}_{q-4})'$, $\bm{\beta}_2 = (1,0, -2.5,0,-1, \bm{0}_{q-4})'$, where $\bm{0}_q$ denotes the $q$-dimensional zero vector. All variables are standardized.

	\item [(ii)] Model 6 (GMM $\bm{x}$, Skewed $y$ given $\bm{x}$): We use the same model as in Model 5 except for $y$: for $h=1, 2$,
	\begin{eqnarray*}
		y = (1,\bm{x}')\bm{\beta}_h + e, ~~~~ \mbox{if}~~ U \in I_h,
	\end{eqnarray*}
	where $e \sim Gamma(1,1)$.
\end{itemize}

We assume the same missing pattern and imputation accuracy metrics as in Simulation One.

As seen from Table \ref{ch5.tb3}, CGMM using the penalized regression method outperforms GMM and PMM in terms of the RMSPE and MAE. The performance of the GMM is worse than the PMM, due to the numerical problems in computing the variance-covariance matrices. The CGMM does not suffer such problems and shows good prediction accuracy.

\begin{table}
	\caption{Average root mean squared prediction error (RMSPE) and mean absolute error (MAE) of three imputation methods based on 1,000 Monte Carlo samples}
	\begin{center}
		\label{ch5.tb3}
		\begin{tabular}{cccc}
			{True Model} & {Method} & {RMSE} & {MAE}\\
			\hline 
			\multirow{3}{*}{Model 5}  &PMM &  {0.8768} & {0.7034} \\ 
			&	GMM & {1.1432} & {0.9384} \\ 
			& CGMM & {0.3912} & {0.2815} \\ 
			\hline
			\multirow{3}{*}{Model 6}  &PMM &  {0.9759} & {0.7697} \\ 
			&	GMM & {1.9265} & {1.7321} \\ 
			& CGMM & {0.6328} & {0.4684} \\
		\end{tabular}
	\end{center}
\end{table}

%Under Model 5, we computed the average ratio of correctly estimated non-zero effects and incorrectly estimated non-zero effects as follows: 	
%\begin{eqnarray*}
%	CR &=& \frac{1}{MC} \sum_{i=1}^{MC} \frac{\mbox{\# correctly estimated non-zero effects }}{\mbox{\# non-zero coefficients}},\\
%	ICR &=& \frac{}{MC} \sum_{i=1}^{MC} \frac{\mbox{ \# incorrectly estimated non-zero effects}}{\mbox{\# zero coefficients}}.
%\end{eqnarray*}  
%As the result, the CR and ICR are computed as 0.966 and 0.057, respectively. 

\section{Application to real data}
\label{sec:app}

We apply the proposed method to the 2017 Korean Household Income and Expenditure Survey (KHIES) conducted by Statistics Korea, which motivates our study. One purpose of the KHIES is to provide an up-to-date information about Korean household welfare-related status. It measures several different types of income items per each person in a household such as earned income, business income, financial income, property income, and other types of incomes as well as expenditure-related items and basic demographic information. Earned income is the primary study variable considered in this study. 

\begin{figure}
	\begin{center}
	\includegraphics[width=5in]{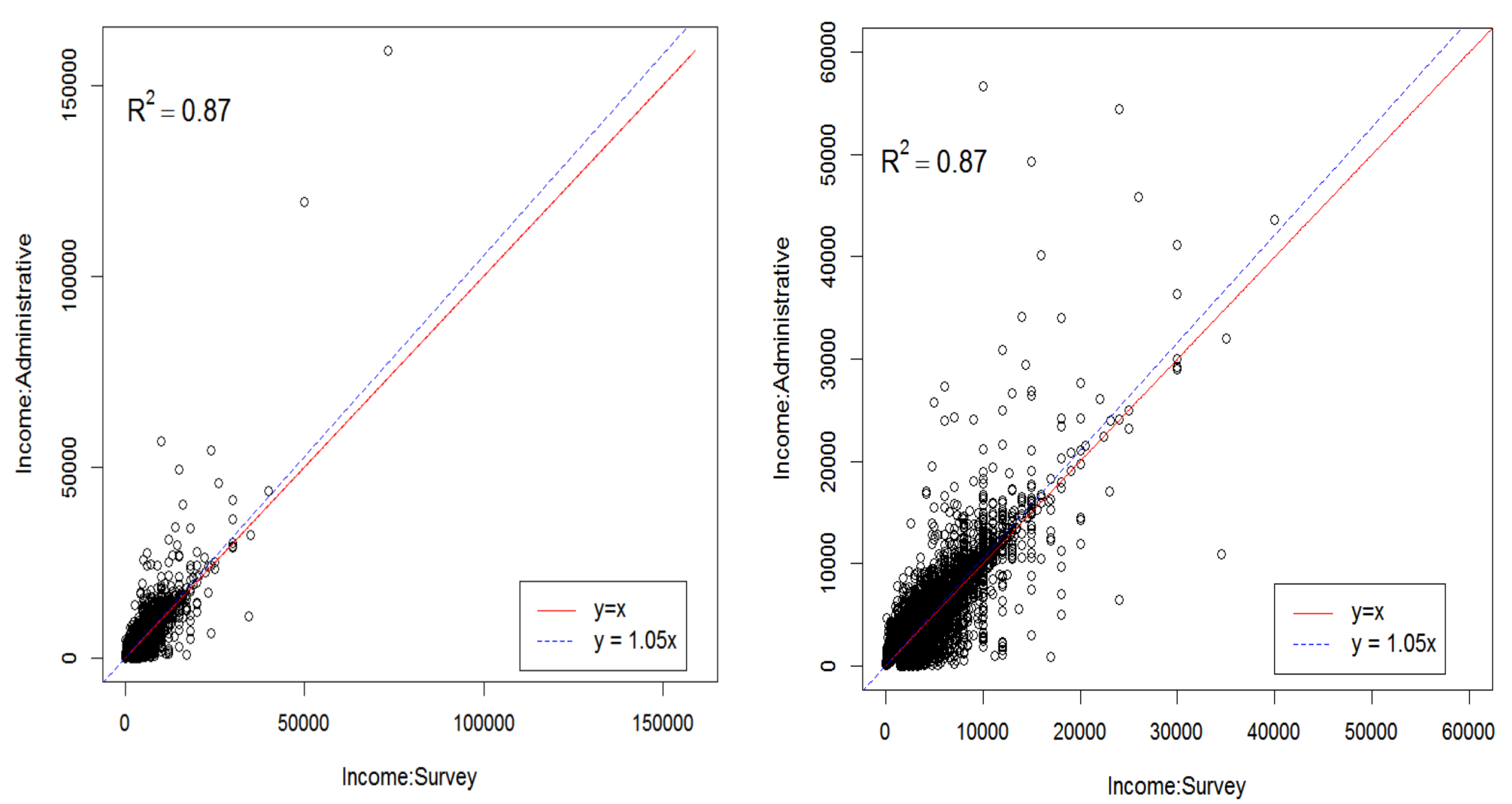}
	\end{center}
	\caption{Scatterplots of the survey and administrative earned incomes for the matched respondents in the KHIES (Unit: KRW 10,000)}
	\label{DA.fig1}
\end{figure}

Since 2014, income tax administrative data has been accessible to Statistics Korea and the accurate information about earned income is available for each person in the sample using personal identification number (PIN). However, some participants in the sample do not reveal PIN. In this case, their tax information about earned income is not available. The overall matching rate of the KHIES sample is about $85\%$. As shown in Table \ref{DA.tb1} and Figure \ref{DA.fig1}, the earned incomes from the two data sources are highly correlated, however, there are still differences, which suggests measurement errors in the reported income in KHIES.

To get improved estimates for some target population quantities, it is desirable to use more reliable administrative records for the matched  respondents in the survey. The challenge is that the administrative data are available only for the matched respondents and there might exist inconsistencies between the matched and unmatched respondents. In this study, we regard unavailable administrative records for the unmatched respondents as item nonresponse and apply the proposed imputation method.

\begin{table}
	\caption{ Summary statistics of survey and administrative annual earned incomes for the matched and unmatched groups (Unit: KRW 1,000)}
	\label{DA.tb1}
	\begin{center}
		\begin{tabular}{cccccc}
			\multicolumn{2}{l}{} & 1st Qu. & Median & Mean & 3rd Qu.\\
			\hline
			\multirow{2}{*}{Matched} & Survey  & 14,400 & 24,000 & 31,450 & 40,000  \\
			{} & Administrative &  12,000 & 22,280 & 31,990 & 42,200\\
			\hline
			\multirow{2}{*}{Unmatched} & Survey & 15,000 & 24,000 & 29,290 & 37,100  \\
			& Administrative & \multicolumn{4}{c}{NA}\\
		\end{tabular}
	\end{center}
\end{table}

%\begin{figure}[h!tb] \centering

%\includegraphics{Images/dc5}

%\isucaption{Durham Centre}
%\label{mgraph}
%\end{figure}

Let $y$ be the study variable of our interest, earned income observed from the administrative data and $\tilde{y}$ be the earned income from the survey data which is subject to some measurement errors. Let $\bm{x}$ be a vector of covariates commonly observed from the two data sets, such as age and education. By matching the survey data to the administrative data, we now have the data structure as in Table \ref{DA.tb2}.

\begin{table}
	%	\centering
	\caption{ Data structure }
	\label{DA.tb2}
	\begin{center}
		\begin{tabular}{rccc}
			& $x$ & $\tilde{y}$ & $y$   \\
			\hline
			Matched & $\checkmark$ & $\checkmark$ & $\checkmark$ \\
			Unmatched & $\checkmark$ & $\checkmark$ &  \\
		\end{tabular}
		\\
		\vspace{0.2cm}
		NOTE: ``$\checkmark$'' implies availability of data.
	\end{center}
\end{table}

In Figure \ref{DA.fig1}, we observe that $\tilde{y}$ and $y$ are highly correlated with increasing variation for large $\tilde{y}$ in which the ratio imputation of $y$ using $\tilde{y}$ only is appealing. To improve the prediction accuracy, we can divide data into several cells so that observations are homogeneous within each cell and then perform ratio imputations within each cell. Such cell-formation can be determined by $\tilde{y}$ and other covariates $\bm{x}$. However, we do not have clear evidence of a relationship between $\tilde{y}$ and $\bm{x}$, and $\tilde{y}$ is very skew-distributed itself. This motivates the following finite mixture model which avoids a direct specification of a joint distribution of $\tilde{y}$ and $\bm{x}$. For $g=1, \ldots, G$,
\begin{eqnarray*}
	\log \left\{ \frac{f_1(\tilde{y}_{i}, \bm{x}_i \mid z_i=g)}{f_1(\tilde{y}_{i}, \bm{x}_{i} \mid z_i=1)}\right\} &=& (1, \tilde{\bm{x}}_{i}') \bm{\alpha}_g, \\
	y_i \mid \tilde{y}_{i}, {x}_i, z_i=g &\sim & N(\tilde{y}_{i}\beta_g, \sigma_g^2),
\end{eqnarray*}
where $\tilde{\bm{x}}_{i}=(\tilde{y}_{i}, \bm{x}_i')'$ and $\bm{\alpha}_1=\bm{0}$. The imputation model is then given by 
\begin{eqnarray*}
	f(y \mid \tilde{y}, \bm{x}) = \sum_{g=1}^G \pi_g(\tilde{y}, \bm{x}) f(y \mid \tilde{y}; \beta_g, \sigma_g^2),
\end{eqnarray*}
where 
\begin{eqnarray}
\pi_g(\tilde{y}, \bm{x}) =  \frac{\exp((1,\tilde{\bm{x}}') \bm{\alpha}_g)}{1+\sum_{k=2}^G \exp((1, \tilde{\bm{x}}')\bm{\alpha}_k)}, \label{pig}
\end{eqnarray}
for $g=2, \ldots, G$, and $\pi_1(\tilde{y}, \bm{x}) = 1- \sum_{g=2}^G \pi_g(\tilde{y}, \bm{x})$.
Let $\hat{\bm{\theta}}$ denote the maximum likelihood estimates and we compute imputed values of $y$ for the unmatched respondents in the survey as 
\begin{eqnarray*}
	\hat{y}_i^\ast = \sum_{g=1}^G \hat{\pi}_g(\tilde{y}_{i}, \bm{x}_i) \tilde{y}_{i}\hat{\bm{\beta}}_g,
\end{eqnarray*} 
which is a weighted sum of cell ratio estimation, where $\hat{\pi}_g(\tilde{y}, \bm{x})$ is $\pi_g(\tilde{y}, \bm{x})$ in (\ref{pig}) evaluated at $\bm{\alpha}=\hat{\bm{\alpha}}$. We consider $G=\{1, \ldots, 10\}$ and then select $G$ minimizing $BIC(G)$. In this data, $G=4$ was selected. 

Table \ref{DA.result1} shows the estimated parameters of the proposed mixture model with $G=4$. It successfully distinguishes a cell in which the survey and administrative earned incomes are exactly same, from other cells, and we can see from the estimated $\bm{\alpha}_g (g=1, \ldots, 4)$ that the survey earned income more contributed to form such cells than age and education. Table \ref{DA.result2} presents that the average imputed earned income is higher than the mean of the survey earned income, which is consistent with the difference between the survey and administrative incomes for the matched respondents. The imputed estimates with $95\%$ confidence intervals for several quantities presented in Table \ref{DA.result3}, where jackknife is used to estimate the variance of the imputed estimates. Based on the confidence intervals, the proposed imputed results show non-negligible differences from the estimates only based on the survey earned income.

\begin{table}
	\caption{Estimated parameters of the proposed mixture model with $G=4$}
	\label{DA.result1}
	\begin{center}
		\begin{tabular}{ccccccc}
			$g$ & $\beta_g$ & $\sigma_g^2$ & $\alpha_{g,0}$ & $\alpha_{g,Age}$ & $\alpha_{g,Edu}$ & $\alpha_{g,Survey}$ \\ 
			\hline
			1 & 1.00 & 0.00 & 0.00 & 0.00 & 0.00 & 0.00 \\ 
			2 & 1.03 & 37.06 & 0.88 & -0.11 & -0.08 & 2.51 \\ 
			3 & 1.44 & 5912.03 & -1.28 & 0.38 & -0.10 & 2.63 \\ 
			4 & 0.96 & 605.17 & 1.49 & -0.23 & -0.05 & 2.25 \\ 
		\end{tabular}
	\end{center}
\end{table}

\begin{table}
	%	\centering
	\caption{Summary statistics of survey and administrative/imputed earned incomes for the matched/unmatched groups (Unit: KRW 1,000)}
	\label{DA.result2}
	\begin{center}
		\begin{tabular}{cccccc}
			\multicolumn{2}{l}{} & 1st Qu. & Median & Mean & 3rd Qu.\\
			\hline
			\multirow{2}{*}{Matched} & Survey  & 14,400 & 24,000 & 31,450 & 40,000  \\
			{} & Administrative &  12,000 & 22,280 & 31,990 & 42,200\\
			\hline
			\multirow{2}{*}{Unmatched} & Survey & 15,000 & 24,000 & 29,290 & 37,100  \\
			& Imputed & 15,130 & 24,310 & 29,720 & 37,610\\
		\end{tabular}
	\end{center}
\end{table}

\begin{table}
	%	\centering
	\caption{Imputation results with $95\%$ confidence interval and estimates of survey earned incomes (Unit: KRW 1,000)}
	\label{DA.result3}
	\begin{center}
		\begin{tabular}{cccc}
			& Survey estimate & Imputed estimate & $95\%$ Confidence Interval \\ 
			\hline
			1st Qu. & 14,450 & 12,104 & (11,904, 12,303) \\ 
			Median & 24,000 & 22,778 & (22,164, 23,391) \\ 
			Mean & 31,204 & 31,675 & (31,213, 32,137) \\ 
			3rd Qu. & 40,000 & 41,396 & (40,592, 42,199) \\
		\end{tabular}
	\end{center}
\end{table}

\section{Concluding remarks}
\label{sec:concluding}

We introduce a new class of more flexible mixture densities than the GMM for semiparametric imputation. In the proposed mixture model, we assume a Gaussian model for the conditional distribution of the study variable given the auxiliary variables, however, the marginal distribution of the auxiliary variables is not necessarily Gaussian. The marginal distribution of the auxiliary variable within each mixture component can be viewed as a density ratio model, which covers the Gaussian model as a special case. As the proposed model uses the mixture model for the conditional distribution directly, the penalized likelihood technique for high dimensional problem is applicable and the prediction accuracy can be greatly improved when the true model is sparse, as demonstrated in the second simulation study. The computation for parameter estimation is relatively easy to implement and fast, as it does not use the MCMC computation in the EM algorithm.

In this study, we assume that the log of density ratio is a linear combination of the auxiliary variables. As an extension, we can consider a more flexible density ratio assumption using a nonparametric kernel method. Also, the proposed method is only applicable to continuous study variables. Developing conditional mixture model for categorical study variable is an important extension. Such extensions will be topics for future research.

\bigskip
\begin{center}
{\large\bf SUPPLEMENTARY MATERIAL}
\end{center}

\begin{description}

\item [~] Supplementary material includes proofs of the theorems and computational details. 

\end{description}

\bibliographystyle{Chicago}

\bibliography{Bibliography-LK}
\end{document}